\documentstyle[twoside,fleqn,espcrc2]{article}

\newcommand{\AmS}{{\protect\the\textfont2
  A\kern-.1667em\lower.5ex\hbox{M}\kern-.125emS}}

\newcommand{\plb}[3]{Phys. Lett. {\bf B#1} (#2) #3} 
\newcommand{\prl}[3]{Phys. Rev. Lett. {\bf #1} (#2) #3}
\newcommand{\prd}[3]{Phys. Rev. {\bf D#1} (#2) #3}
\newcommand{\npb}[3]{Nucl. Phys. {\bf B#1} (#2) #3}
\newcommand{\npbps}[3]{Nucl. Phys. {\bf B}(Proc. Suppl.) {\bf #1} (#2) #3}

\title{
\hfill
\parbox{3cm}{\normalsize KUNS-1530 
}\\
Symmetry and Symmetry Restoration 
of Lattice Chiral Fermion in the Overlap Formalism
}

\author{Y. Kikukawa\address{Department of Physics, Kyoto University, 
Kyoto 606-8502, Japan}%
        \thanks{
This talk is based on the collaborations with H.~Neuberger (Rutgers Univ.), 
A.~Yamada (Univ. of Tokyo)  and T.~Aoyama (Kyoto Univ.).
This work is supported in part by Grant-in-Aid
for Scientific Research from Ministry of Education, Science
and Culture(\#10740116,\#10140214).}
}
\begin{document}

\begin{abstract}
Three aspects of symmetry structure of lattice chiral fermion 
in the overlap formalism are discussed. 
By the weak coupling expansion of the overlap Dirac operator, 
the axial anomaly associated to the chiral 
transformation proposed by L\"uscher is evaluated and is shown
to have the correct form of the topological charge density for
perturbative backgrounds. 
Next we discuss the exponential suppression of the self-energy 
correction of the lightest mode in the domain-wall fermion/truncated 
overlap. 
Finally, we consider a supersymmetric 
extension of the overlap formula in the case of the chiral multiplet 
and examine the symmetry structure of the action.
\end{abstract}

\maketitle

\section{Introduction}

The overlap formalism of the chiral determinant \cite{overlap} 
provides a well-defined lattice regularization of chiral determinant. 
It can reproduce the known features of the chiral determinant 
in the continuum theory.
This suggests that the overlap formalism is a promising 
building block for the construction of lattice chiral gauge models.

When applied to vector-like theories like QCD, the overlap
formalism also provides a satisfactory description of massless
Dirac fermion. It has been shown by Neuberger
\cite{overlap-Dirac-operator} that the overlaps for 
the massless Dirac fermion can be written as a single determinant 
of a Dirac operator, 
which explicit form is defined by 
\begin{equation}
\label{overlap-dirac-operator}
a D
= 1 + X \frac{1}{\sqrt{X^\dagger X}} 
= 1 + \gamma_5 \frac{H}{\sqrt{H^2}} ,
\end{equation}
where $X$ is the Wilson-Dirac operator, 
\begin{eqnarray*}
X &=& 
\left\{ \gamma_\mu 
\frac{1}{2}\left( \nabla_\mu- \nabla_\mu^\dagger \right)
+ \frac{a}{2} \nabla_\mu \nabla_\mu^\dagger 
- \frac{1}{a} m_0 
\right\}
\end{eqnarray*}
$( 0 < m_0 < 2)$ and $H=\gamma_5 X$. 
The remarkable point about this Dirac operator is 
that it satisfies the Ginsparg-Wilson 
relation \cite{ginsparg-wilson-rel,GW-fixed-point-D,GW-overlap-D}.
It implies that the symmetry breaking is restricted to 
unphysical local terms. This is the clue to escape the 
Nielsen-Ninomiya theorem. Moreover, 
as shown by L\"uscher \cite{exact-chiral-symmetry}, 
the fermion action has exact symmetry under the lattice 
chiral transformation.
For the flavor-singlet chiral transformation,
the functional measure is not invariant and causes the anomaly. 
This anomaly is given exactly by the index of the overlap
Dirac operator \cite{index-finite-lattice,exact-chiral-symmetry}. 
\[ 
- a {\rm Tr} \gamma_5 D 
= 2 N_f \, {\rm index}(D) 
= - N_f {\rm Tr} \left( \frac{H}{\sqrt{H^2}} \right)  .
\]
In the overlap formalism, it is measured
by the spectrum asymmetry of the Hamiltonian $H$, 
which has been considered as the topological charge 
by Narayanan and Neuberger \cite{overlap}.

\section{Weak coupling expansion of overlap Dirac operator}

In the weak coupling expansion, we have calculated 
the anomaly explicitly and have shown that it has the 
correct form of the topological charge density 
for perturbative background \cite{weak-coupling-expansion-overlap-D}.

It is straightforward to obtain the weak coupling expansion 
of the overlap Dirac operator, Eq.~(\ref{overlap-dirac-operator}). 
For the explicit expressions of the expansion, the author 
refers the reader to \cite{weak-coupling-expansion-overlap-D}.
At the second order, it contains the following term
\begin{eqnarray*}
&& \int \frac{d^4 k}{(2\pi)^4}
    {
            \frac{\left[\omega(q)+\omega(k)+\omega(p)\right]}
            {\omega(q)\omega(k)\omega(p)} 
}\times 
\nonumber\\
&& \,
\frac{X_0(q) X_1^\dagger(q,k) X_0(k) X_1^\dagger(k,p) X_0(p)}
{\left[\omega(q)+\omega(p)\right]
             \left[\omega(q)+\omega(k)\right]
             \left[\omega(k)+\omega(p)\right]} ,
\end{eqnarray*}
where 
\[
\omega(p)= \sqrt{
 \sin^2 p_\mu + \left( \sum_\mu(1-\cos p_\mu) - m_0 \right)^2}.
\]
The Wilson-Dirac operator $X$ is expanded up to the second 
orders in the momentum space as 
$X=X_0 + X_1 + X_2 + {\cal O}(g^3)$ .
Only this term has more than three gamma matrices, which could
contribute to the anomaly. This term has a factor symmetric 
in $\omega$'s. If we recall the following identity, 
\begin{eqnarray*}
&& 
\frac{1}{2} 
\frac{\omega_1+\omega_2+\omega_3}
                 {\left(\omega_1+\omega_2\right)
                  \left(\omega_2+\omega_3\right)
                  \left(\omega_1+\omega_3\right) \omega_1 \omega_2
                  \omega_3}
 \nonumber\\
&& 
=
\int_{-\infty}^\infty \frac{d\omega}{2\pi} 
\frac{1}
{ \left(\omega^2+\omega_1^2\right)  
  \left(\omega^2+\omega_2^2\right)  
  \left(\omega^2+\omega_3^2\right) } ,
\end{eqnarray*}
this structure suggests that
the coefficient of the anomaly can be expressed in terms 
of a five-dimensional propagator.
\[
X_5(\omega,p) 
\equiv  i \gamma_5 \omega + X_0(p). 
\]
In fact, the vertex function of the anomaly 
can be expanded in terms of the external 
momenta without encountering any IR divergence. 
The coefficient of the leading operator of dimension four can be 
expressed as a winding number of a map defined 
by $X_5(\omega,k)$, which is a function from
$T^4 \times R$ to $S^5$.
Then we see that the correct axial anomaly 
is reproduced in the covariant form:
\begin{eqnarray}
&&\lim_{a\rightarrow 0}
\left(- a \sum_n {\rm tr} \left\{ \alpha_n \gamma_5 D_{nn} \right\} 
\right)
\nonumber\\
&&=  \frac{g^2}{32 \pi^2} \, N_f \, 
\int d^4 x \, \alpha(x) \,
\epsilon_{\lambda\mu\sigma\nu} \, 
F^a_{\lambda\mu}(x) F^a_{\sigma\nu}(x) .  \nonumber
\end{eqnarray}
The relation between this axial anomaly and the covariant anomaly 
discussed in the context of the overlap formalism in 
\cite{overlap,consistent-and-covariant-anomaly,geometrical-aspect-of-anomaly} 
will be discussed in more detail elsewhere.

\section{
Exponentially suppressed self-energy correction
in the domain-wall fermion / truncated overlap }

The next topic is about the exponentially suppressed self-energy
correction in the domain-wall fermion / truncated overlap.
The Shamir's variant of the domain-wall fermion can be regarded 
as a collection of a finite number, say $N$, of Wilson-Dirac
fermions. As discussed by Neuberger, it becomes more clear if we 
perform the chirally-asymmetric parity transformation in the flavor space.
In this basis, the mass matrix becomes hermitian (N=4):
\begin{eqnarray*}
&& M^H_{st}(n,m) = M_{st}(n,m) P =P M^\dagger_{st}(n,m)  \\
\nonumber\\
&=&
\left( \begin{array}{cccc}
              0 & 0 & -1 & B+M_0 \\
              0 & -1 & B+M_0 & 0 \\
              -1 & B+M_0 & 0 & 0 \\
              B+M_0 & 0 & 0 & 0 \end{array} \right) ,
\end{eqnarray*}
where $P$ stands for the parity operator acting on the flavor space.

At the tree level, this hermitian mass matrix is diagonalized 
in the momentum space by the Tchebycheff polynomials, 
$u_s(x) =\frac{ \sin s \omega }{ \sin \omega}$ $(x=\cos \omega)$
supplemented by the boundary condition
\[  \frac{ u_{N+1}(x)}{u_N(x)}=  
  \frac{\sin (N+1)\omega}{\sin N \omega} 
= \frac{1}{ \sum_\nu (1-\cos p_\nu)+M_0} .\]
Then, the mass eigenvalues are given by the following formula:
\[
  m(p) = \frac{1}{u_{N+1}(x)}=\frac{\sin \omega }{\sin (N+1) \omega} .
\]
For $\frac{1}{\sum_\nu (1-\cos p_\nu)+M_0}  < 1 + \frac{1}{N} $, 
a single mode with an imaginary $\omega$ appears and it gives 
an exponentially small eigenvalue,
\[
  m_0(p) =\frac{\sinh \lambda }{\sinh (N+1) \lambda} \simeq 
       \exp ( - N \lambda ) .
\]

The question here is that this exponentially small 
mass eigenvalue survives the radiative correction due 
to the gauge field or not.
Since there is no reason based on symmetry at finite
flavor $N$, we need to check it explicitly, at least in 
perturbation theory. 
Aoki and Taniguchi have calculated the self-energy correction
at one-loop for asymptotically large flavor 
$N$ \cite{exponential-suppression-domain-wall}.  
Neuberger has examined the underlying mechanism for generating 
almost massless fermion \cite{almost-massless-fermion}.  
We have shown that the calculation at one-loop
can be done keeping the number of flavor $N$ finite 
in the diagonal basis of the leading mass 
matrix \cite{exponential-suppression-truncated-overlap}.
Thus we can check explicitly the exponential suppression 
of the self-energy correction to the lightest mode at one-loop.
For more detail of the calculation at finite $N$, 
the author refers the reader 
to \cite{exponential-suppression-truncated-overlap}.

\section{Chiral symmetry and supersymmetry in the overlap formalism}

Finally, we discuss a supersymmetric extension of the overlap 
formalism in the case of the free chiral 
multiplet \cite{chiral-multiplet-in-overlap}.
In the free theory, the overlap Dirac operator may be written 
in the following form. 
\[ D= \gamma_\mu \nabla_\mu^X + M^X , \]
where 
\begin{eqnarray*}
\nabla_\mu^X 
&=& \frac{1}{2}
\left( \partial_\mu - \partial_\mu^\dagger\right)
\frac{1}{\sqrt{X_0^\dagger X_0}} , \\
M^X &=& 1 + \frac{1}{\sqrt{X_0^\dagger X_0}} 
\left( \frac{1}{2} \partial_\mu \partial_\mu^\dagger - m_0 \right) .
\end{eqnarray*}
These difference operators satisfy the following identity, 
which is another expression of the Ginsparg-Wilson relation:
\[ 2 M^X = - \left( \nabla_\mu^X\right)^2 + (M^X)^2 . \]

Using these kinetic and mass operators, we can write 
a local action of the two-component Weyl fermion and 
the complex boson, which possesses manifest supersymmetry 
as follows:
\begin{eqnarray*}
&& \sum_n \left\{ 
\bar \psi_n \sigma_\mu \nabla_\mu^X \, \psi_n  
- \phi_n^\ast \left(\nabla_\mu^X\right)^2  \phi_n - F_n^\ast F_n 
\right\}
\nonumber\\
&&
+\sum_n 
\frac{1}{2} \left\{
\psi_n^T \epsilon M^X \psi_n 
- \bar \psi_n \epsilon M^X \bar \psi_n^T  \right\}
\nonumber\\
&&
+ \sum_n \left\{ \phi_n M^X F_n + F_n^\ast M^X \phi_n^\ast \right\} .
\end{eqnarray*}
In addition to the manifest supersymmetry, 
the fermionic action possesses the chiral symmetry under the 
following transformation in terms of the two-component Weyl 
fermion field,
\begin{eqnarray*}
  \delta \psi_n &=& - \psi_n 
+ \frac{1}{2} \left( M^X \psi_n - \sigma_\mu^\dagger \nabla_\mu^X
  \epsilon \bar \psi_n^T \right) , \\
  \delta \bar \psi_n &=& + \bar \psi_n 
- \frac{1}{2} \left( \bar \psi_n M^X + \psi_n^T \epsilon
  \sigma_\mu^\dagger \nabla_\mu^X\right) . 
\end{eqnarray*}
Moreover, the bosonic action has two $U(1)$ symmetries. One of them 
is the bosonic extension of the lattice chiral symmetry of L\"uscher:
\begin{eqnarray*}
  \delta \phi_n &=& \phi_n 
- \frac{1}{2} \left( M^X \phi_n - F_n^\ast \right) , \\
  \delta F_n &=& F_n 
- \frac{1}{2} \left( M^X F_n - \left(\nabla_\mu^X \right)^2
  \phi_n^\ast \right) .
\end{eqnarray*}

This result implies that all the symmetries of the target continuum 
theory of the free chiral multiplet can be manifest on the lattice 
in the overlap formalism.
The introduction of the supersymmetric cubic interaction seems 
hard, because of the breakdown of the Leibniz rule on the lattice.
We need further considerations in this direction.


\begin{thebibliography}{9}
\bibitem{overlap}
R.~Narayanan and H.~Neuberger, 
\npb{412}{1994}{574};
\prl{71}{1993}{3251};
\npb{443}{1995}{305}.

\bibitem{overlap-Dirac-operator}
H.~Neuberger, \plb{417}{1998}{141}.

\bibitem{ginsparg-wilson-rel}
P.~H.~Ginsparg and K. G. Wilson, \prd{25}{1982}{2649}. 

\bibitem{GW-fixed-point-D}
P.~Hasenfratz, \npbps{63}{98}{53}.

\bibitem{GW-overlap-D}
H.~Neuberger, Phys. Lett. {\bf B427} (1998) 353.

\bibitem{exact-chiral-symmetry}
M. L\"uscher, Phys. Lett. {\bf B428} (1998) 342.

\bibitem{index-finite-lattice}
P.~Hasenfrats, V.~Laliena and F.~Niedermayer, 
Phys. Lett. {\bf B427} (1998) 125.

\bibitem{weak-coupling-expansion-overlap-D}
Y.~Kikukawa and A.~Yamada, \\
{\tt hep-lat/9806013}. 

\bibitem{consistent-and-covariant-anomaly}
S.~Randjbar-Daemi and J.~Strathdee, 
Phys. Lett. {\bf B402} (1997) 134.

\bibitem{geometrical-aspect-of-anomaly}
H.~Neuberger, 
{\tt hep-lat/9802033}. 

\bibitem{exponential-suppression-domain-wall}
S.~Aoki and Y.~Taniguchi, {\tt hep-lat/9711004};
Nucl. Phys. {\bf B}(Proc. Suppl.){\bf 63} (1998) 290.

\bibitem{almost-massless-fermion}
H.~Neuberger, Phys. Rev. {\bf D57} (1998) 5417.

\bibitem{exponential-suppression-truncated-overlap}
Y.~Kikukawa, H.~Neuberger and A.~Yamada, 
Nucl. Phys. {\bf B526} (1998) 572.

\bibitem{chiral-multiplet-in-overlap}
T.~Aoyama and Y.~Kikukawa, 
to appear in Phys. Rev. D, {\tt hep-lat/9803016}.

\end{thebibliography}
\end{document}